\documentclass[aps,prl,reprint,superscriptaddress,]{revtex4-1}
\usepackage{amsmath, amsthm, amssymb}
\usepackage{graphicx}

\usepackage{epstopdf}
\usepackage{hyperref}

\begin{document}


\title{Controlling plasmonic Bloch modes on periodic nanostructures}



\author{B. Gjonaj}
\email[]{b.gjonaj@ee.technion.ac.il}
\altaffiliation{Current address:Technion Israel Institute of Technology}


\affiliation{FOM-Institute for Atomic and Molecular Physics AMOLF,
Science Park 104, 1098 XG Amsterdam, The Netherlands}

\author{J. Aulbach}
\affiliation{FOM-Institute for Atomic and Molecular Physics AMOLF,
Science Park 104, 1098 XG Amsterdam, The Netherlands}

\author{P. M. Johnson}
\affiliation{FOM-Institute for Atomic and Molecular Physics AMOLF,
Science Park 104, 1098 XG Amsterdam, The Netherlands}

\author{A. P. Mosk}
\affiliation{Complex Photonic Systems, Faculty of Science and
Technology, and MESA+ Institute for Nanotechnology, University of
Twente, PO Box 217, 7500 AE Enschede, The Netherlands}

\author{L. Kuipers}
\affiliation{FOM-Institute for Atomic and Molecular Physics AMOLF,
Science Park 104, 1098 XG Amsterdam, The Netherlands}
\author{A. Lagendijk}
\affiliation{FOM-Institute for Atomic and Molecular Physics AMOLF,
Science Park 104, 1098 XG Amsterdam, The Netherlands}


 \begin{abstract}
We study and actively control the coherent properties of Surface
Plasmon Polaritons (SPPs) optically exited on a nano-hole array.
Amplitude and phase of the optical excitation are externally
controlled via a digital spatial light modulator (SLM) and SPP
interference fringe patterns are observed with high contrast. Our
interferometric observations revel SPPs dressed with the Bloch modes
of the periodic nano-structure. The momentum associated with these
Dressed Plasmons (DP) is highly dependent on the grating period and
fully matches our theoretical predictions. We show that the momentum
of DP waves can in principle exceed the SPP momentum. Actively
controlling DP waves via programmable phase patterns offers the
potential for high field confinement applicable in sensing, Surface
Enhanced Raman Scattering and plasmonic structured illumination
microscopy.
\end{abstract}
\maketitle

Important systems such as biological cells, single molecules, and
nanodevices, strongly interact with visible light on sub-wavelength
scales. Yet standard microscopy and related applications in sensing
and imaging are diffraction limited. Plasmonics
\cite{barnes_surface_2003} offers an alternative route to control
light with sub-wavelength precision through the excitation of
Surface Plasmon Polaritons (SPPs)
\cite{ozbay_plasmonics:_2006,polman_applied_2008}. These surface
waves, bound to a metal dielectric interface, are a hybrid mode of
photons and electronic change-density oscillations. The intrinsic
momentum associated with these evanescent waves is higher than that
of free propagating photons. Thus, for a fixed light frequency, SPPs
have a higher effective refractive index and tighter confinement of
electromagnetic energy \cite{schuller_plasmonics_2010}.

Innovation in nano fabrication has enabled a remarkable degree of
control over SPPs using metallic nanostructures. Specially tailored
samples allow plasmonic waves to be coupled into the topological
modes of a fabricated structure. The light field confinement, and
therefore the resolution, addressed through the mode volume of these
geometrically Dressed Plasmons (DP) exceeds by more than one order
of magnitude that of standard SPP confinement. Successful geometries
include coupled nanoantennas
\cite{Schuck2005,Muhlschlegel10062005,novotny2011} that fully
localize modes in the gap between neighboring antennas, and
V-grooved \cite{sondergaard2010,stockman2004} and nanowire
\cite{krenn2002,evold2009} waveguides that support 1D propagating
modes deeply confined inside the waveguide.

Yet there are limitations in using dressed plasmons for sensing
applications. The electromagnetic field is only locally enhanced due
to the fixed geometry of the structure yielding very high resolution
but no field of view. Furthermore the specimen has to be inserted
within the few nanometers width of the waveguide or the gap between
nanoantennas, an extremely difficult task using current methods.
However, theoretical works \cite{sentenac_subdiffraction_2008,
bartal_subwavelength_2009} have shown that it is possible to use
periodic nanostructures \cite{Zheludev2011}, such as well designed
gratings \cite{Rodrigo2010}, to support extended DP waves to obtain
both high resolution and large field of view. Furthermore, actively
controlling these DP waves has the potential for plasmonic
structured illumination microscopy \cite{wei2010} and related
applications in imaging and sensing.

Here we show experimental observation and control of extended
dressed plasmons supported by periodic nanostructures. Using a
Spatial Light Modulator we shape the amplitude profile of the
incident laser beam over a large 2D field of view. The SLM is imaged
onto the surface of the sample thus addressing each pixel of the SLM
to a corresponding area on the sample.This arrangement allows us to
measure with high contrast fringe patterns generated from two
counterpropagating SPP waves. Tuning the SLM phase pattern allows
these fringes to be shifted and/or tilted at will. The momentum
associated with the standing waves shows strong dependence on the
lattice period of the grating and revels the Bloch-mode dressing of
the surface plasmons. Combining high momentum DP with focusing and
scanning experiments \cite{gjonaj_active_2011} has the potential to
revolutionize far field bio-sensing applications.

\begin{figure}[t!]\
\centering
\includegraphics[width=0.45\textwidth]{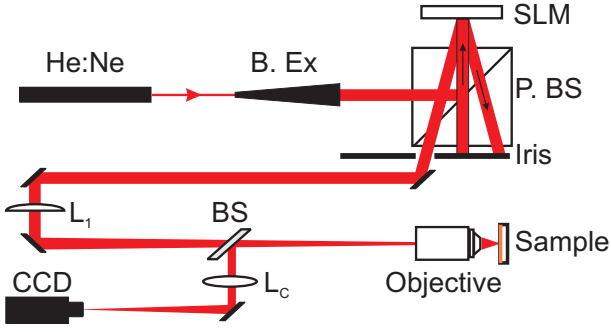}\\
\caption{Experimental Setup. The Spatial Light Modulator (SLM) is
projected onto the sample via the imaging system (lens $L_1$,
objective). The sample is imaged on the CCD camera via the  system
(objective, lens $L_C$). The abbreviations B. Ex, BS and P. BS stand
for beam expander, beam splitter and polarizing beam splitter
respectively. Inset: SEM image (5~x~5 $\mu$m) of 450 nm nanohole
array.} \label{Figure_setup}
\end{figure}

A diagram of the setup is given in Fig.~\ref{Figure_setup}. The SLM is
imaged on the sample via a lens (L1) and the objective, referred to
hereafter together as the imaging system. The SLM is at the focal
plane of lens $L_1$ (focal length 130 cm). The image at infinity
created by $L_1$ is projected onto the sample at the focal plane of
the objective. Our SLM (Holoeye LC-R 720) is a reflective display
based on Twisted Nematic Liquid Crystal on Silicon technology. The
display has a total of 1280 x 768 pixels operating at 60 Hz with a
response time of 3 ms. Each pixel is 20 $\mu$m in size and addressed
with a 8-bit voltage. The objective (Nikon LU PLAN FLUOR P 100X) is
infinity corrected and metallurgic (no coverslip compensation) with
a Numerical Aperture (NA) of 0.9 and a magnification of 100 times
(defined for a tube lens of 20 cm focal length). The focal length of
$L_1$ is 6.5 times larger than that of the standard tube lens
yielding a corresponding 650 times demagnification of the image. The
distance between $L_1$ and the objective is 1 m i.e. smaller than
the focal length of $L_1$ (non-telecentric imaging system). In this
configuration the average angle of illumination is position
dependent, which is an
important condition for the SPPs launching.

The light emitted in reflection from the sample is imaged on the CCD
(AVT Dolphin F145 B) using lens $L_C$ as tube lens. This light
includes both the direct reflection of the illuminating beam and the
scattered light from SPPs. Thus the resulting image is a combination
of both the SLM amplitude pattern and the generated SPP pattern. To
distinguish between the two we choose illumination patterns that
allow SPP observation in a non-illuminated area. The amplitude and
phase of the excitation pattern is controlled by applying the
4-pixel technique \cite{vanPuttenSLM08} to the SLM. Four adjacent
pixels are grouped into a superpixel by selecting a first
diffractive order with the neighbor-pixel fields being $\pi$/4 out
of phase. In this work we use 32 x 32 superpixels. Every SLM
superpixel is imaged on a sample area of 440~x~440 nm$^2$ containing
nearly one unit cell of the grating. Such a superpixel grouping
provides continuous modulation over full amplitude ($A\in[0 ,1]$)
and phase ($\Phi\in[0
,2\pi]$) ranges with a cross modulation of less than 1\%.\\

Our samples, nanohole arrays similar to those used typically for
Enhanced Optical Transmission experiments, were fabricated using
focused ion beam milling. A 200 nm gold film was deposited on top of
1 mm BK7 glass substrate with a 2 nm chromium adhesion layer. Square
holes were milled with sides of 177 nm. The hole array covers an
area of 30 x 30 ${\mu}$m$^{2}$. Five samples were fabricated with
array periods ($a_0$) varying from 350 nm to 450 nm.  The sample was
placed with the gold side towards the objective to observe SPP waves
from the gold-air interface. We calculate the SPP momentum for
incident radiation of $\lambda_0$~=~633~nm ($k_0$~=~
2$\pi$/$\lambda_0$) using tabulated values
\cite{johnson_optical_1972} of the dielectric constants of gold
$\varepsilon_m$ and air $\varepsilon_d$

\begin{equation}\label{equation1}\\
k_{S}= k_0
{\rm{Re}\sqrt{\frac{\varepsilon_m\varepsilon_d}{\varepsilon_m+\varepsilon_d}}}={(m,n)}k_{G}+k_{0}\sin\theta,
\end{equation}

where the last equality expresses the fact that the SPP momentum is
a vectorial sum of the ($m$,$n$)$k_G$ grating orders
($k_G=2\pi/a_0$) and the in-plane component of the incident light.
With our oblique illumination scheme, the average angle of incidence
$\theta$ is not uniform but position dependent.

\begin{figure}[b!]\
\centering
\includegraphics[width=0.45\textwidth]{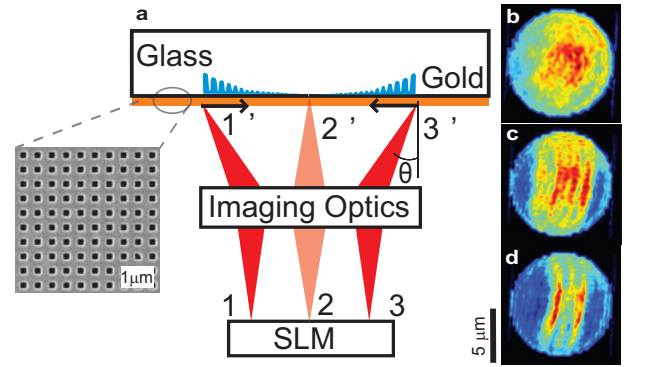}\\
\caption{(a) Sketch of the sample illumination. Each SLM point is
imaged on the surface of the sample with a different average angle
of incidence $\theta$. (b-d) Reflection from different samples
illuminated with a uniform amplitude and phase profile. (b) On the
bare gold sample the reflection is also uniform since no SPPs are
launched. (c) The dark areas of low reflection from the 400 nm hole
array indicate the angular (spatial) bands for SPP launching. (d)
These bands are sample dependent as shown for the 425 nm hole array.
Inset: SEM image of the samples.  } \label{sample}
\end{figure}

This illumination scheme and its role on how SPPs are launched is
illustrated in Fig.~\ref{sample}. Each SLM's superpixel is projected onto
the sample with a different average angle of incidence
(Fig.~\ref{sample}a) and thus with a different in-plane component of the
incident light. The momentum conservation described in
Eq.~\ref{equation1} will be satisfied only within specific angular bands
which are position dependent. In Fig.~\ref{sample}b-d we show the surface
of three different samples illuminated with a uniform amplitude
profile across the SLM with horizontal polarization.

For the reference bare gold film and a uniform SLM amplitude and
phase profile, the reflected image is identical to incident beam
profile since no SPP can be launched (Fig.~\ref{sample}b). When the same
uniform amplitude and phase profile is projected onto a nanohole
array, dark and bright areas are clearly distinguishable as shown in
Fig.~\ref{sample}c-d. Dark areas correspond to suppressed reflection from
the sample. We interpret these dark areas as the spatial (angular)
bands that satisfy Eq.~\ref{equation1} and thus where plasmons are
efficiently excited from the incident light. The location of these
bands strongly depends on the array momentum. Even a 25 nm variation
of the array period from $a_0$ = 425 nm (Fig.~\ref{sample}c) to $a_0$ =
400 nm (Fig.~\ref{sample}d) yields a spatial band shift of nearly
2~$\mu$m.

SPPs waves launched in the momentum matched bands propagate towards
each other and interfere (Fig.~\ref{sample}). Yet this interference
pattern is observed on a high background due to the direct
reflection of the incident light. To remove the background and
enhance the contrast of the SPP interference pattern we spatially
design the incident amplitude profile with ``on" areas of amplitude
A = 1 and an ``off" background of A = 0. Each ``on" area is composed
of 10~x~8 superpixels and is located in the vicinity of the two
symmetric angular bands. The SPP interference patterns are then
observed in the central non-illuminated area which is our SPP field
of view.

\begin{figure}[b!]\
\centering   \includegraphics[width=0.45\textwidth]{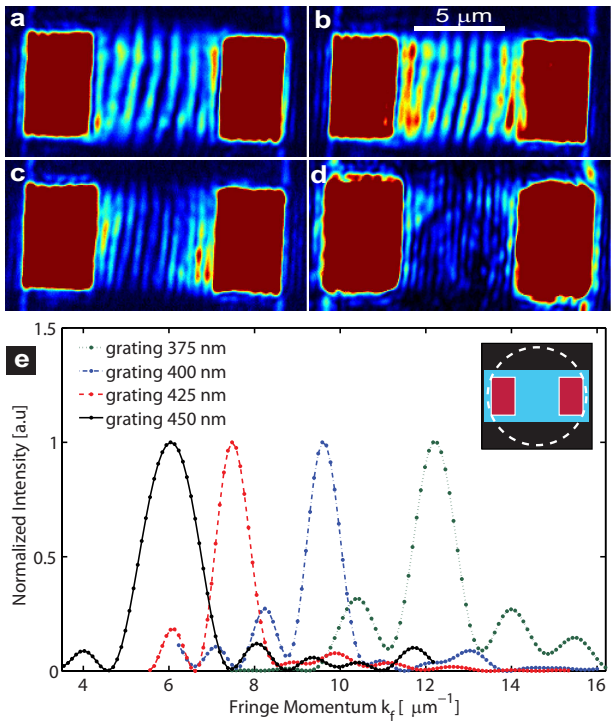}\\
\caption{SPP fringe formation via counter propagating waves. The
image geometry and the incident amplitude profile are shown in the
inset. The polarization of the incident light is horizontal. (a-d)
we observe different fringe patterns for array periods of 375 nm
(a), 400 nm (b), 425 nm (c) and 450 nm (d). In (e) are shown the
line Fourier transforms of these fringe patterns. }
\label{amplitudes}
\end{figure}

Results from this designed amplitude profile are shown in
Fig.~\ref{amplitudes}. When the two counterpropagating SPP waves launched
in the ``on" areas interfere, a standing wave pattern of intensity
is created. For SPPs propagating on an ideally smooth and
non-corrugated sample we expect the period of the fringe pattern to
be half the SPP wavelength ($\lambda_S=2\pi/k_S=590 nm$). Instead,
the measured fringe period is found to be sample dependent
(Fig.~\ref{amplitudes}a-d). We measured fringe periods $P$ of
$1\pm0.05~\mu$m, $0.85\pm0.05~\mu$m, $0.65\pm0.05~\mu$m,
$0.5\pm0.05~\mu$m and $0.45\pm0.05~\mu$m for grating pitches of 450
nm, 425 nm, 400 nm, 375 nm and 350 nm respectively. The different
filling fractions of our samples, that perturb the SPP wavelength
within few percent, can not explain the large deviations we observe.

\begin{figure}[t!]
\includegraphics[width=0.45\textwidth]{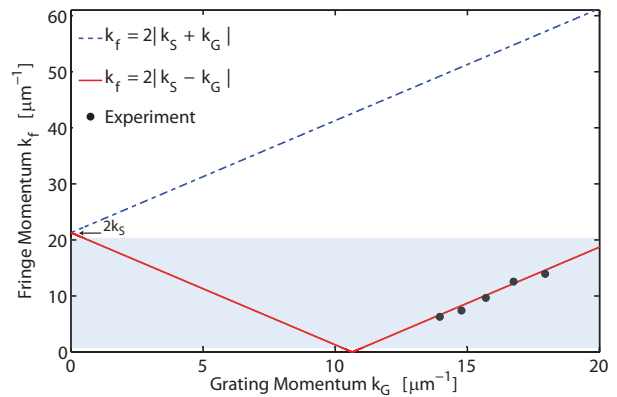}\\
\caption{SSP fringe momentum versus the grating momentum. The
experimental data shows SPPs convoluted with the $m=-1$ Bloch mode
of the arrays. The other Bloch modes can not be resolved due to
limited detection bandwidth (light blue square).} \label{keyvectors}
\end{figure}

We attribute the fringe patterns to a mixing of the original SPP
wave with the hole array \cite{lienau2003}. We can analyze the
results using a one dimensional model because for all our samples we
observe only horizontal propagation. Theoretically there are two
ways to mix SPPs with the hole array: intensity mixing (expected for
incoherent forms of scattering such as fluorescence) and field
convolution (expected for coherent scattering processes). We will
discuss both ways even though the experimental observations confirm
only the field convolution. We first consider intensity convolution:
a SPP standing intensity pattern with momentum $2k_S$ is formed, but
since we observe the pattern through the scattering of a periodic
structure with momentum $k_G$, the fringe momentum appears to be
$2k_S \pm k_G$. This intensity convolution does not match the
experimental observations. The situation is completely different for
the field convolution: the hybridization of the bare SPPs with the
Bloch modes of the array results in dressed plasmonic (DP) waves of
momentum $k_S + m\cdot k_G$ ($m$ integer). These DP waves then
result in standing intensity patterns of momentum $2(k_S +m\cdot
k_G)$.

A comparison between experiment and the amplitude convolution
approach for these DP waves is shown in Fig.~\ref{keyvectors}. The
modulus of the fringe momentum ($k_f$) is plotted against the module
of the grating momentum ($k_G$). The two lines are the theoretical
predictions for SPPs convolved with the first positive ($m=1$) and
the first negative ($m=-1$) grating orders. The experimental data
perfectly follow only the $m=-1$ curve. The first positive order is
not observed in our far field measurement due to its evanescent
non-radiative nature and the limited resolution of our setup. The
distribution of fringe momenta can be expressed except for a
normalization factor as
\begin{equation}\label{equation2}
P_f(k)=B(k)\sum_{m\in\mathbb{Z}}{\eta_
m\cdot\delta\left(k-2\left|k_S+mk_G\right|\right)},
\end{equation}
where every delta represents the standing pattern from one of the
$m$ orders of the array, $\eta_ m$ represents the coupling
efficiency of SPPs into this $m^{-th}$ order and $B(k)$ is the
momentum bandwidth of our detection optics. Our bandwidth is shown
as the light blue quadrate in Fig.~\ref{keyvectors} and we approximate it
with a step function limited by the optical diffraction limit and
the SPP field of view (the distance between the two ``on" areas).
Upon inserting this bandwidth in Eq.~\ref{equation2} only SPP
hybridization with the $m=-1$ term survives because all other DP
modes have fringe momentum that exceeds the diffraction limit.

\begin{figure}[h!!!!]
\includegraphics[width=0.45\textwidth]{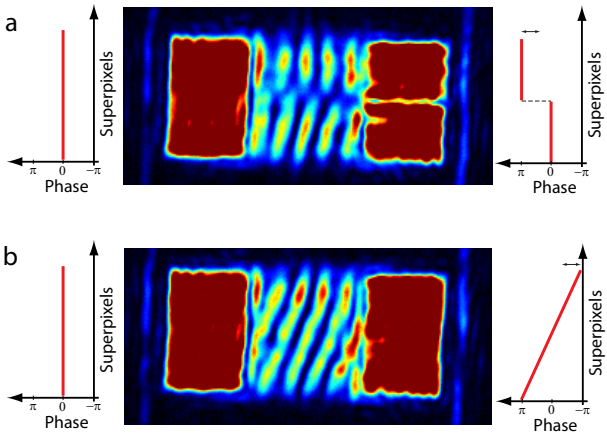}\\
\caption{ Scanning the fringes by phase tuning. (a) The phase of the
left ``on" area is kept constant while the phase of the right ``on"
area has a step jump between the lower and upper part as shown in
the wings of the figure. The resulting fringe pattern in the upper
part is shifted compared to the lower one. (b) The fringes are
rotated due to a linearly incrementing phase on the right ``on"
area.} \label{phaseScan}
\end{figure}

We can scan the fringe pattern across the sample by varying the
phase delay between the two ``on" areas and thus introducing an
optical retardation that will translate the DP fringes. We
experimentally prove this phase scanning principle for the $m=-1$ DP
modes as shown in Fig.~\ref{phaseScan}a where the upper half of the
right``on" area is out of phase with the rest of the illuminated
areas. The different phase delays result in a translated fringe
pattern in the upper part. The line scan resolution (fringe
translation) is given by our digital phase control: 256 steps from 0
to $2\pi$ phase delay. In alternative, by applying a linear phase
difference between the two ``on" areas, the standing pattern will
result in tilted plasmonic fringes (angular scan) as shown in
Fig.~\ref{phaseScan}b.

The predicted presence of the $m=1$ DP mode, which represents a sub
100 nm period intensity beating on top of the observed fringe
pattern, combined with the our ability to scan the pattern across
the sample, suggest interesting prospects for subwavelength imaging.
Due to the diffraction limit we can not resolve this fast beating in
the current setup. However it should be possible, using near field
imaging, to calibrate this sub 100 nm intensity pattern for
different fringe patterns (line and angular scans). Once calibrated,
the sample surface could be used to image sub 100 nm objects with
only far field probing and image correlations.

We have shown here the observation of Bloch-mode dressed surface
plasmon polaritons (DP) propagating on nanohole arrays of different
subwavelength periodicities. We recorded the standing intensity
pattern of two counterpropagating DP waves. The dependence of the
measured fringe period  on the period of the nano structure is
perfectly described by a simple model of plasmonic Bloch mode
interference. By actively imposing well programmed phase relations
to these plasmonic Bloch modes we achieved full control of their
interference fringe patterns. Bloch dressed SPPs are 2D propagating
waves that can achieve high momentum and thus actively controlling
their interference patterns has potential for super-resolution
biosensing and imaging applications.

\section{Acknowledgments}
We thank Hans Zeijlermaker and Dimitry Lamers for sample
fabrication. This work is part of the research program of the
``Stichting voor Fundamenteel Onderzoek der Materie", which is
Financially supported by the ``Nederlandse Organisatie voor
Wetenschappelijk Onderzoek".

\bibliography{References}

\end{document}